\documentstyle[manuscript, epsf, aps]{revtex}

\begin{document}
\title{Nonadiabatic conditional geometric phase shift with NMR }
\author{Wang Xiang-Bin\footnote{correspondence author, email: wang@qci.jst.go.jp}
 \hskip 0.3cm and  \hskip 0.2cm Matsumoto Keiji\footnote{email:keiji@qci.jst.go.jp}\\
         Imai Quantum Computation and Information Project,\\ ERATO,
Japan Sci. and Tech. Corp.\\Dani Hongo White Bldg. 201, 5-28-3, 
Hongo Bunkyo, Tokyo 113-0033, Japan}
\maketitle 
\begin{abstract}{ A conditional geometric phase shift gate, which is fault tolerant 
to certain types of errors due to its geometric nature, was realized recently
 via nuclear magnetic resonance(NMR)  
under adiabatic conditions. 
By the adiabatic requirement, the result is inexact unless the Hamiltonian
 changes  extremely slowly. However, in quantum computation, 
everything must be completed within the decoherence
time. High running speed of every gate in quantum computation is demanded because the 
power of a quantum computer can be exponentially proportional to the maximum  number of 
logic gate operations that can be made sequentially within the decoherence time.  
The adiabatic condition makes any fast conditional Berry phase(cyclic adiabatic geometric phase) shift gate impossible. 
Here we show that by using a newly designed sequence of simple operations with
an additional vertical magnetic field,  the conditional geometric phase 
shift gate can be run nonadiabatically. Therefore geometric quantum computation can be done at the same rate as usual quantum computation.}  
\end{abstract}

A fault-tolerant quantum logic gate\cite{knill1} is the central issue in realizing  the basic
constituents of a quantum information processor. 
Quantum computation via the controlled geometric phase shift\cite{ekert} provides 
a nice scenario to this purpose.
Due to its geometric property, 
a geometric phase\cite{berry,aha} shift can be 
robust with respect to certain types of operational errors. In particular, 
suppose that the spin(qubit) undergoes  a random 
fluctuation about its path in the evolution. Then
the final value of the geometric phase shift 
will not be affected provided that the random fluctuation does not change
the total area\cite{ekert}. 

Recently, it was reported\cite{ekert,nature,quant,zoller,aver,los} that the conditional 
Berry phase( adiabatic cyclic geometric phase) shift can be used in  quantum computation. 
In particular, in ref\cite{ekert},
 an experiment was done
with NMR\cite{cory,gers,jones,jones1,book} under the adiabatic condition.   

Demanding on both running speed and precision of every  gate in a quantum computer 
is quite high. It has been reported recently that\cite{ekert}, due to the 
requirement of 
adiabatic condition,
 faster running speed 
causes severe distortions to the results\cite{ekert}. The adiabatic condition
makes the fast speed and high precision conflict each other in the 
conditional Berry phase shift gate.
  Increasing the running speed of the logic gate can exponentially increase the power
of a quantum computation such as quantum factorization\cite{shor}. 
It seems that the geometric phase shift gate will  only be   practical if
it can run at a speed comparable to that of the usual quantum gate.
Therefore one is tempted to find an easy  way to make the
 conditional geometrical phase shift
through  {\em non}adiabatic state evolution. In this letter, we present an
easy  scheme to make the geometric phase shift $non$adiabatically
with NMR. We start from the exact control of the state evolution on a cone.\\
{\bf Exact state evolution  on the cone.}
It is well known that a spin half nucleus can gain a geometric
phase shift in the conical evolution.
We now assume that the initial spin state is on the cone and we will demonstrate a method to 
 make it evolve nonadiabatically.
The Hamiltonian for a spin  in a constantly rotating magnetic field
\footnote{We shall extend it to the case of arbitrary time-dependent rotating speed in the end of this section.}  
is
\begin{eqnarray}\label{e1}
H(t)=\left[\omega_0\sigma_z +\omega_1 \sigma_x(t)\right]/2
.\end{eqnarray}
Here $\sigma_x(t)=\left(\begin{array}{cc}
0&e^{-i\gamma t}\\e^{i\gamma t}
&0\end{array}\right)$, 
$\omega_0$ is the amplitude of vertical field and $\omega_1$ is the amplitude of the horizontal field, which
is rotating around $z-$axis in  the constant  angular speed $\gamma$.
The initial state  $|\psi_0\rangle$ is an
eigenstate of $H_0=H(0)$. 
Explicitly,  $|\psi_0\rangle=\cos\frac{\theta}{2}|\uparrow\rangle+\sin\frac{\theta}{2}|\downarrow\rangle$,
$\cos\theta=\omega_0/\sqrt{\omega_0^2+\omega_1^2}$.
To know the cause of the state distortion, we study the time evolution first.(The similar idea has been used
for the adiabatic rotational splitting with NQR previously\cite{robert}.)
The time evolution operator $U$ generated by the Hamiltonian $H_0$ is determined by
the time-dependent Schrodinger equation
\begin{eqnarray}
i\frac{\partial}{\partial t}U_0=H(t) U_0.
\end{eqnarray}
Solving this equation we obtain
\begin{eqnarray}
U_0(t)=e^{-i\gamma\sigma_z t}e^{-i(H_0-\gamma \sigma_z/2)t}.
\end{eqnarray}
In particular, at time $t=\tau=2\pi/\gamma$, when the external field completes a $2\pi$ rotation, the state is 
evolved to
\begin{eqnarray}
\psi(\tau)=e^{-i\pi-i(H_0-\gamma \sigma_z/2)\tau}|\psi_0\rangle.
\end{eqnarray}
The adiabatic approximation assumes that
at time $\tau$, the state completes a cyclic evolution provided that $\gamma$ is small. 
However, due to the decoherence time limitation, $\gamma$ cannot be too small. 
The non-zero $\gamma$ distorts the state in the evolution and 
makes it {\em noncyclic} at time $\tau$ .   
This non-zero $\gamma$ in the adiabatic approximation causes two types of errors. One is that
the noncyclic completion of the evolution at time $\tau$ will cause further errors in the succeeding operations in the sequence.
(A sequence of operations is used 
to remove the dynamic phase in realizing the geometric quantum gate\cite{ekert}.)
The other is that the geometric phase acquired over period $\tau$ is not $-\pi(1-\cos\theta)$, 
as for the ideal adiabatic cyclic evolution. 

Now we give an easy way to exactly control 
 the state evolution on the cone.  
Here the external field can be  rotated around $z-$axis arbitrarily fast. We use $|\psi_0\rangle$,
the eigenstate of $H_0$ for the initial state. 
We  switch on a static vertical magnetic field 
$\omega_z=\gamma$ while the external field is rotated around $z-$axis.
In this way, the new time-dependent Hamiltonian is 
$H_W(t)=\frac{1}{2}\left[(\omega_0+\gamma) \sigma_z+\omega_1 \sigma_x(t)\right]$. The 
time evolution operator $U_W(t)$ generated by this Hamiltonian 
satisfies the Schrodinger equation
\begin{eqnarray}\label{sch}
i\frac{\partial}{\partial t}U_W(t)=H_W(t)U_W(t)
\end{eqnarray}
with the boundary
condition $U_W(0)=1$. The state  at time $t$ is related to the state
at time $0$ by $|\psi(t)\rangle=U_W(t)|\psi_0\rangle$, $\psi_0=\psi(0)$.
 Denoting $R=e^{i \gamma t\sigma_z /2}$ we obtain the following equivalent equation
\begin{eqnarray}
i\partial(RU_W)/\partial t=H_0(RU_W).
\end{eqnarray}
Noting that $H_0$ is time-independent therefore we have $RU_W(t)=e^{-iH_0t}$. 
This is equivalent to
\begin{eqnarray}\label{zzz}
U_W(t)=e^{-i\gamma t\sigma_z/2}e^{-iH_0 t}.
\end{eqnarray}
  Consequently, at any time $t$, state $|\psi(t)\rangle=e^{-i\lambda t}
e^{-i\gamma t\sigma_z /2 }|\psi_0\rangle$ is
exactly the instantaneous eigenstate of Hamiltonian $H(t)$, where
$\lambda$ is the eigenvalue of $H_0$ for eigenstate $|\psi_0\rangle$.
 In particular, at time $\tau$, $|\psi(\tau)\rangle=e^{-i\pi-i\lambda \tau}|\psi_0\rangle$. 
It   only differs to $|\psi_0\rangle$ by a phase factor.
Note that the time evolution operator $U_W(t)$ here is generated by the Hamiltonian 
$H_W(t)$ instead of $H(t)$. 

Here
the additional vertical field $\gamma$ plays an important role. Without this field, the time
evolution operator generated by $H(t)$ is
\begin{eqnarray}
U(t)=e^{-i\gamma t\sigma_z/2}e^{-iH_1 t}
\end{eqnarray}
and $H_1=H_0-\gamma\sigma_z/2$. 
Obviously, this time evolution operator $U(t)$ will distort a qubit with the initial
state $|\psi_0\rangle$ in the evolution. But this $U(t)$ can exactly
control the qubit with the initial state
$|\psi_1\rangle$, the eigenstate of $H_1$. If initially we set the
spin state to $|\psi_1\rangle$, then the spin will evolve exactly on its cone without the 
additional field
$\gamma$.  So,
we have two ways to control the spin evolution exactly. 
We can use the additional field $\gamma$, if the initial state is set to be $|\psi_0\rangle$.
Alternatively, we can also set the initial state to be $|\psi_1\rangle$ and 
then we need not add
any additional field when the external field is rotated.
In this letter, we adopt the former one, i.e. we set the initial spin state to $|\psi_0\rangle$.
\footnote{Our main motivation is to give the nonadiabatic
 quantum gate in the most general case, i.e., we shall
allow the time-dependent rotation of horizontal field $\omega_1$. For this purpose, only the method
using  additional vertical field will work.}

We have assumed above that the field rotates at a constant speed.
Actually, we can easily modify the  above scheme for 
arbitrary time-dependent rotating speed $\gamma_a(t)$, 
 $\int^\tau_0 \gamma_a(t)dt=2\pi$. In this case, we need only change the term $\gamma t$
in $H_W(t)$, $U_W(t)$, $R$ and $\psi(t)$ into $\int^t_0\gamma_a(t')dt'$ accordingly.
Consequently, the the additional vertical field is now a time-dependent 
field $\omega_z(t)=\gamma_a(t)$ instead of a static field. This extension is important
in case it is difficult to rotate the field (or the fictitious field\cite{nature}) in a
 constant angular speed. Punctual results can be obtained
here through the exact feedback system where the value of additional vertical field is always
instantaneously equal to the  angular velocity of the rotating field. 

Thus we see, by adding an additional magnetic field that is equal to the rotating
frequency of the external field, we do get the {\it exact} result. 
 Obviously if we
rotate the field  inversely, additional vertical field  should be 
in the inverse direction($-z$) accordingly. 

{\bf NMR system and the rotational framework.} 
Consider the interacting nucleus spin pair(spin $a$ and spin $b$) in the NMR 
quantum computation\cite{cory,gers,jones,jones1}.
If there is no horizontal field the Hamiltonian for the two qubit system is
$H_i=\frac{1}{2}(\omega_{a} \sigma_{za}+\omega_b\sigma_{zb}+ J \sigma_{za}\cdot\sigma_{zb}) $,
where $\omega_{a}(\omega_b)$ is the resonance frequency for spin $a(b)$ 
 in a very strong 
static magnetic field(e.g. $\omega_a$ can be $500$MHz\cite{ekert}) ,  $J$ is the interacting constant between
nuclei and $\sigma_{za}=\sigma_{zb}=\sigma_z=\left(\begin{array}{cc}
1&0\\0&-1\end{array}\right)$.  
After adding a circularly polarized RF field in horizontal
plane,
the Hamiltonian for spin $a$ in static framework is
\begin{eqnarray}
H'=\frac{1}{2}\omega_0\sigma_z+\frac{1}{2}\omega_a'\sigma_z+\frac{1}{2}\omega_1
\left(\begin{array}{cc}0&e^{-i\omega_a't}\\e^{i\omega_a't}&0\end{array}\right).
\end{eqnarray}
Here $\omega_a'$ and $\omega_1$ are the angular frequency and amplitude of the RF field respectively,
$\omega_0=\omega_a-\omega_a'\pm J$ and the $"\pm"$  sign in front of $J$ is dependent on 
the specific state of spin $b$, up or down respectively. To selectively manipulate spin $a$, we shall 
use the rotational framework that is rotating around $z-$axis
in speed $\omega_a'$.
We assume $\omega_a'$  close to $\omega_a$ but obviously different from $\omega_b$.
The Hamiltonian for spin $a$ in the rotational framework is   
\begin{eqnarray}
H_a=R'H'R'^{-1}+i(\partial R'/\partial t)R'^{-1}=H_0
\end{eqnarray}
and $R'=e^{i\omega_a'\sigma_zt/2}$. Note here $H_0$ is dependent on  state of spin $b$ 
through $\omega_0$.
In the rotational framework, if the horizontal field is rotated in the  angular speed
$\gamma$, 
the Hamiltonian for spin $a$ is just $H(t)$, as defined in eq(\ref{e1}). In the NMR system, we require $|\psi_0\rangle$ 
to be the eigenstate
of $H_0$ in rotational the framework.  Previously\cite{ekert}, state 
$|\psi_0\rangle$ for spin $a$ was produced  adiabatically. 
In the following we demonstrate how to 
nonadiabatically produce the state $|\psi_0\rangle$.
\\{\bf Creating the conditional initial state with NMR.}
In rotating the state around $z-$axis, we have started from state $|\psi_0\rangle$, 
which is the eigenstate
of $H_0$. State $|\psi_0\rangle$ must be created from the spin state $|\uparrow\rangle$ or $|\downarrow\rangle$. 
We need  a $non$adiabatic way to  
create state $|\psi_0\rangle$ for spin $a$.  
Note that we are working in the rotational framework. 
We denote 
$\delta=\omega_a-\omega_a'$.
We can use the following sequence of operations
to create the conditional angle $\theta$.(For simplicity we will call the following sequence as $S$ operation later on.)
We have $S=$
\begin{eqnarray}\label{scheme}
\left[\frac{\pi}{2}\right]^y\rightarrow
J'(\varphi_{\pm}(t_c))
\rightarrow\left[-\delta \cdot t_c\right]^z
\rightarrow \left[\frac{\pi}{2}\right]^x
\rightarrow\left[{-\varphi'}
\right]^y
\end{eqnarray}
Here all terms inside the $[\cdots]$ represent the Bloch sphere
rotation angles caused by  RF pulses. 
The superscripts indicate the axis the Bloch sphere is rotated around. $J'( \varphi_{\pm}(t_c))$ 
is the time evolution over period $t_c$ by the 
Hamiltonian $\frac{1}{2}(\delta\pm J)\sigma_z$.
 This evolution rotates spin $a$ around $z-$axis for an angle
 $\varphi_{\pm}=(\delta\pm J)t_c$.
(Or $\varphi_{\pm}+\pi$, if spin $a$ is down initially. 
For clarity we omit this case and alway assume spin $a$ is initially up.) 
After this $S$ operation, the angle between spin $a$ and the $z$-axis is
$\theta_{\pm}=\frac{\pi}{2}-(\varphi'+\varphi_{\pm})$(see Fig. 1). 
To ensure the state to be eigenstate of Hamiltonian $H_0$
after $S$ operation 
we require $ \tan(\varphi'+Jt_c)=\frac{\delta+J}{\omega_1}$ and $\tan (\varphi'-Jt_c/2)=\frac{\delta-J}{\omega_1}$ simultaneously. This is equivalent to 
\begin{eqnarray}\label{const}
\left\{\begin{array}{c}Jt_c=
(\arctan\frac{\delta+J}{\omega_1}-\arctan\frac{\delta-J}{\omega_1})/2\\
\varphi'=(\arctan\frac{\delta+J}{\omega_1}+\arctan\frac{\delta-J}{\omega_1})/2\end{array}
\right. .\end{eqnarray}
From this constraint, given specific values of $\delta$, 
$J$ and $\omega_1$, we can easily obtain the scaled time $J\cdot t_c$ (Fig. 2)
and the angle $\varphi'$(Fig. 3) in controlling the $S$ operation. In particular, for $\delta/J=1.058$,  the value adopted
in the recent NMR experiment, the relation curves for $Jt_c$ vs $\omega_1$ and $\varphi'$ vs 
$\omega_1$ are shown 
in Fig. 4. \\ 
{\bf The nonadiabatic conditional geometric phase shift.} After the $S$ operation, the state of spin $a$ 
is an eigenstate of Hamiltonian
$H_0$ no matter whether spin $b$ is up or down. 
After this $S$ operation we can use the additional magnetic field  to nonadiabatically
control  the state evolution on the cone. 
However, after the state completes a loop, the total phase shift includes both dynamic and geometric 
contribution\cite{piller}. It is possible for the state to evolve on the dynamic phase path if we  choose certain specific value of the rotating speed
$\gamma$( and also the additional field $\gamma$). For this task, we require
\begin{eqnarray}\label{restr}
\gamma=-\frac{\omega_1^2+\omega_0^2}{\omega_0}.
\end{eqnarray}
The negative sign in the right hand side indicates that tha additional field $\gamma$
is anti-parallel with the field $\omega_0$.
By this setting, the total magnetic field in $H_1$ is always 
"perpendicular" to the state vector expressed in the Block ball. 
One can easily show that the instantaneous dynamical phase
\begin{eqnarray}
<\psi_0|U_W^\dagger(t)H(t)U_W(t)|\psi_0>=0.
\end{eqnarray}

In case that
$\gamma$ is time dependent, to remove the 
dynamical phase,
we can use the time-dependent magnetic field $\omega_1(t)$ and $\omega_0(t)$.
We require
\begin{eqnarray}
\gamma(t)=\frac{\omega_1^2(t)+\omega_0^2(t)}{\omega_1(t)}
\end{eqnarray}
and
\begin{eqnarray}
\omega_0(t)/\omega_1(t)=\omega_0(0)/\omega_1(0)=\tan\theta
\end{eqnarray}

The two qubit case is a bit different from the single qubit case.
Qubit $b$ could be either up or down. We need choose the appropriate $\gamma$ value so that
the instantaneous dynamic phase for qubit $a$ is always zero no matter qubit $b$ is up or down.
For this purpose, besides equation(\ref{restr}), we require
\begin{eqnarray}
\omega_1=\sqrt{\delta^2-J^2}.
\end{eqnarray} 
With this setting
 we propose the following 
scheme
to make the $non$adiabatic conditional geometric phase shift.    
\begin{eqnarray}
S\rightarrow\left(\begin{array}{c}\omega_z\\{C}\end{array}\right)
\rightarrow S^{-1}
.\end{eqnarray}
The term $\left(\begin{array}{c} {\omega_z}\\{ C}\end{array}\right)$  
 represents  
doing operation $C$  with an additional
vertical field $\gamma$. 
$C$ represents
 rotating the external field around $z-$axis for $2\pi$ in a uniform speed $\gamma$. 
We should choose the rotating direction so that additional field $\gamma$
is anti-parallel the field $\delta\pm J$, ($\delta>J$). 
The  dynamical phase shift after the 
operations is $0$.
So the scheme raised here can remove the dynamic phase and
retain only the geometric phase. 
It can be shown that the geometric phase
acquired after the total sequence of operations is $\Gamma_+$, $\Gamma_-$, $-\Gamma_-$ and 
$-\Gamma_+$ respectively for the four
different initial state
($|\uparrow\uparrow\rangle,|\uparrow\downarrow\rangle,
|\downarrow\uparrow\rangle,|\downarrow\downarrow\rangle$), 
$
\Gamma_\pm=-\pi\mp 2\pi \cos\theta_\pm
,$
$\cos\theta_\pm=\frac{\delta\pm J}
{\sqrt{(\delta\pm J)^2+\omega_1^2}}=\sqrt{\frac{\delta\pm J}{2\delta}}$.

{\bf Concluding remark.} To make the conditional geometric phase shift gate 
we need be able to control two 
elementary operations. One is to exactly control the cyclic state evolution
on a cone; the other is to produce the initial state on the cone with an angle 
$\theta_{\pm}$ conditional on the other bit. 
Previously\cite{ekert}, these two elementary operations were done  
adiabatically. 
We have shown that both of these two tasks can be done nonadiabatically.
We have designed a new scheme for the nonadiabatic
conditional geometric phase shift
gate. This  makes it possible to run the geometric quattum gate in a speed
 comparable to
that of the normal quantum gate.
The idea on nonadiabatic geometric phase shift gate demonstrated by the NMR system here
should in principle  also work for the other two level systems, 
such as Josephson Junction system\cite{nature} and the harmonic oscillator system\cite{quant,wang}, 
where the decoherence time can be much shorter.

{\bf Acknowledgment} We thank Prof Imai Hiroshi for supporting,
Dr Hachimori M., Mr Tokunaga Y and Miss Moriyama S for various helps. 
W.X.B thanks Prof. Ekert A(Oxford) for suggestions. W.X.B. also thanks Dr. Pan J.W.(U. Vienna), Dr Kwek LC(NIE) for discussions
 and Prof
Oh CH(NUS) for pointing out ref\cite{nature}.

{\it Note added:} This manuscript is  a revised version of Ref.\cite{lett1}
with corrections made according to Ref.\cite{lett2}. In this version,
we removed all unpublished contents in the old versions therefore the new version includes only the contents published in\cite{lett1,lett2}. For an extended version with all details, one may refer to quant-ph/0108111\cite{8111}, which was posted to the arXiv one year ago. The original scheme
appeared in Ref.\cite{lett1} does not filter the dynamical phase. This
was first pointed out by Tim Piller et al in August 2001. We then replaced
the old multiloop scheme by a new single loop scheme with dynamical phase free
path. This correction was first presented as the version 4 of this manuscript 
and also as quant-ph/0108111 in August 2001, and finalized in 2002\cite{lett2}. 
  
\begin{figure}
\begin{center}\epsffile{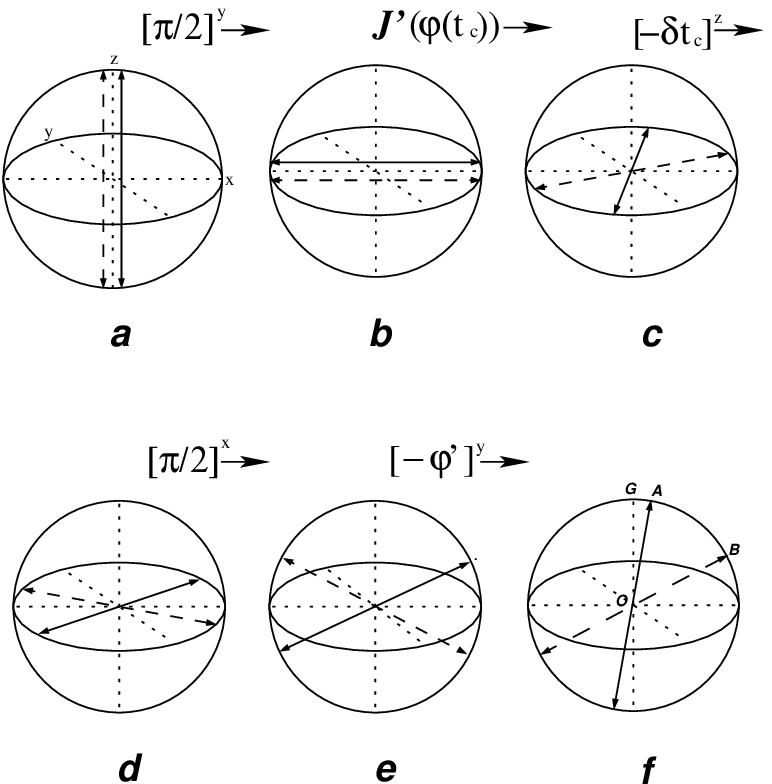}
\caption{ {\bf Using $S$ operation to produce 
the eigenstate of $H_0$ nonadiabatically.} Picture   {\sl a} shows the possible initial state
for spin $a$. The up(down) arrow on the solid line represents the up(down) state for spin $a$
while spin $b$ is up. 
The up(down) arrow on the dashed line represents the up(down) state for spin $a$
while spin $b$ is down. In picture {\sl f},  
angle $GOA$ is $\theta_+ =\arctan \frac{\omega_1}{\delta +J} $ and
angle $ GOB$ is $\theta_-=\arctan \frac{\omega_1}{\delta -J} $.
 }\end{center}
    \end{figure}
\begin{figure}
\begin{center}
\epsffile{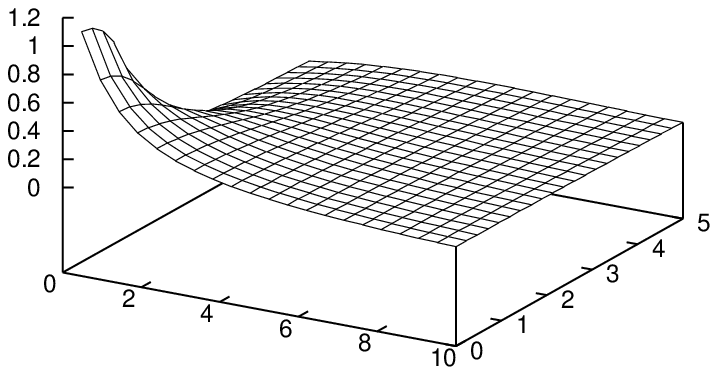}
\end{center}
\caption{{\bf The time control in $S$ operation}. 
Parameter $\omega_1/J$ varies from 0 to $10$, $\delta/J$ varies from $0$ to $5$. 
Vertical axis is for the  scaled time $Jt_c$.}
\end{figure}
\begin{figure}
\begin{center}
\epsffile{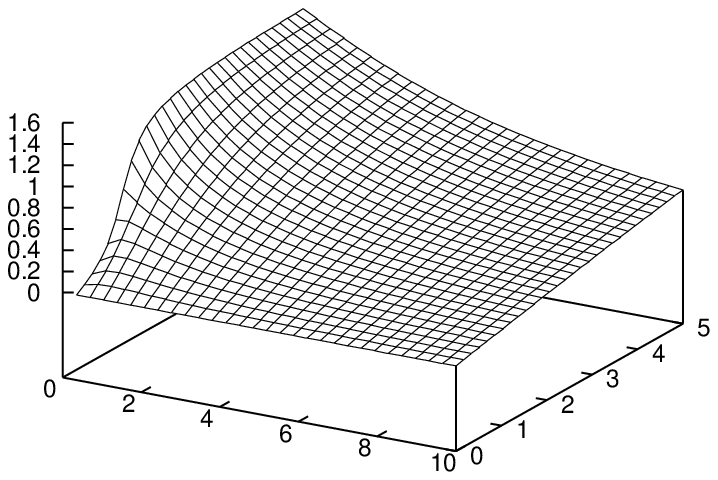}
\end{center}
\caption{{\bf The rotating angle control in $S$ operation}. 
Parameter $\omega_1/J$ varies from $0$ to $10$, $\delta/J$ varies from $0$ to $5$. 
Vertical axis is for 
the angle $\varphi'$. }
\end{figure}
\begin{figure}
\begin{center}
\epsffile{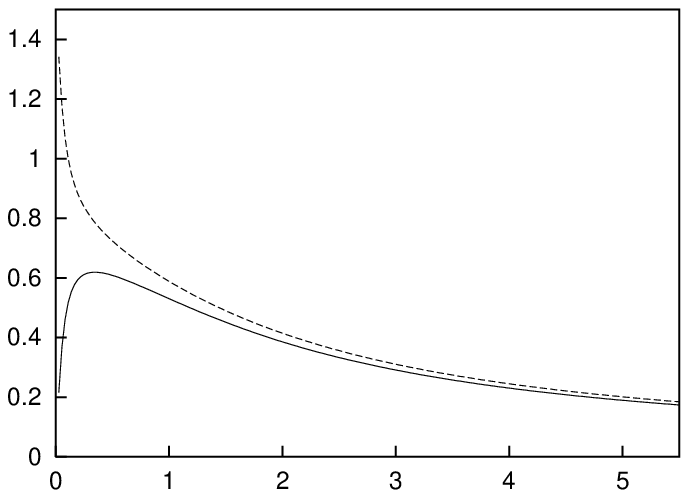}
\end{center}
\caption{ {\bf The time and rotating angle control in $S$ operation for specific $\delta/J$ value}. 
Horizontal axis represents $\omega_1/J$. 
Vertical axis is for the scaled control $Jt_c$(the solid line) or
the angle $\varphi'$(dashed line). Here $\delta/J=1.058$, as for the experimental condition[2].
By this figure, given the specific values of $\omega_1/J$, we can always find the corresponding
point in the two curves thus we can take the suitable time control(for $t_c$) and rotation
control(for $\varphi'$) in the $S$ operation.} 
\end{figure}

\end{document}